# Star Formation Studies at the VLT[†]

Claude Bertout [1], Bo Reipurth [2], Fabien Malbet [3]

[1]Observatoire de Grenoble, Université Joseph Fourier, BP 53, 38041
Grenoble cedex 9, France
[2]European Southern Observatory, Casilla 19001, Santiago 19, Chile
[3]Jet Propulsion Laboratory, Pasadena, USA

**Abstract:** We present an overview of what can be done in the field of star formation using the VLT and its advanced instrumentation, and especially with the very high angular resolution provided by the VLT in the VLTI mode. Herbig-Haro objects and early stellar evolution, pre-main sequence binaries, circumstellar disks and surface properties of young stars are investigated. We show that the VLT and VISA will be not only very useful to better understand star formation processes, but indeed necessary in order to advance in certain subjects.

## 1 Introduction

Instead of trying to cover the entire field of star formation, we focus here on some areas where the gain in angular resolution brought about by the VLT and VLTI will bring decisive progress in our understanding of the physical processes that govern the formation of stars. We shall therefore discuss mainly the circumstellar environment of low-mass young stellar objects.

Over the last ten years, considerable progress about the nature of these objects has been achieved. Comprehensive surveys of molecular clouds uncovered several phases in the formation of stars, ranging from embedded protostars that are studied best at millimeter and submillimeter wavelengths to pre-main sequence, optically visible stars with various degrees of activity. Among these, one now distinguishes between (a) weak-emission line T Tauri stars (WTTSs), which are magnetically active pre-main sequence stars with late spectral type that display strong, variable X-ray flux, flare activity, and large, cool spots; (b) classical T Tauri stars (CTTSs), which are thought to be WTTSs surrounded by accretion disks with accretion rates in the range $10^{-8} - 10^{-6}$ M$_\odot$/yr. The disk produces IR excess and interacts with the underlying star to produce the observed blue and UV radiative excess and to drive observed collimated jets and winds.

---

[†] Invited talk, Proceedings of the ESO Workshop "Science with the VLT", eds. Danziger & J., Walsh, J., Garching, June 1994.

Some open questions currently being investigated are the following: Why are there two classes of young low-mass stars? What is the origin of magnetic field (fossil or dynamo-generated)? What is the nature of the angular momentum transport mecanism in the disk? What is the nature of star/disk interaction (boundary layer vs. accretion column)? What mechanism drives the jet and what is the nature of the observed wind/disk connection?

In addition to the two abovementioned types of young stars, there is a third. much more rare class, the FU Orionis stars, which are erupting CTTSs with apparent mass-accretion rates of up to $10^{-4}$ $M_\odot$/yr. FU Ori stars drive the optical jets discussed below as well as massive molecular outflows. The frequency of outbursts, their nature, and the nature of the jet driving engine are topics of active current research. It should from the outset be pointed out that star formation studies are among the most rapidly evolving subjects in contemporary astrophysics, and it is likely that some of the problems we outline here will have changed character by the time the VLT becomes operational.

## 2 Herbig-Haro Jets and Early Stellar Evolution

About 10 years ago it was recognized that some Herbig-Haro objects (which are nebulous patches commonly found in star-forming regions) take the form of highly collimated jets. In the intervening years such HH jets have been subject to intense study (for a recent review, see Reipurth & Heathcote 1993). Unlike the better known extragalactic jets, HH jets emit line radiation, they are so close that they are on the verge of being resolvable from the ground, and their kinematics can be explored through proper motion and radial velocity studies. Thus, it is possible to derive detailed physical properties for these objects, so that HH jets have emerged as Rosetta stones for the study of highly collimated flows and the earliest stages of stellar evolution.

In most cases HH jets are ejected from very young stars still deeply embedded in their parental clouds. When the extinction is not excessive one can sometimes observe two oppositely directed lobes, each with typical dimensions of 1 to 2 arcminutes, or roughly 0.2 pc. Such jets consist of a highly collimated chain of faint knots, with widths of 0.6″ to 0.8″ and separations of several arcseconds, terminating in a working surface where the flow rams into the ambient medium. The characteristic line spectra from HH flows have been successfully modelled as emission from shocks, and shock models mapping extensive grids of parameter-space have given important insights into the flow physics. Radial velocities and proper motions of HH jets yield typical space velocities of the order of a few hundred km/sec, suggesting dynamical ages of a thousand years or so.

It is commonly found that HH jets have multiple working surfaces along their flow axes. This is best understood in terms of episodic outbursts of the driving source, and HH jets have therefore been linked to FU Orionis eruptions (e.g. Reipurth 1989). Such events are interpreted as massive accretion episodes in the circumstellar disks of T Tauri stars. Although the precise formation mechanism of HH jets is still not known, they may be linked to the disposal of infalling high angular momentum material, probably through interaction with magnetic fields. As the jets stream away from the newborn stars, they interact with the ambient cloud material, and may actually drive the molecular flows commonly observed around young embedded stars (e.g. Raga & Cabrit 1993).

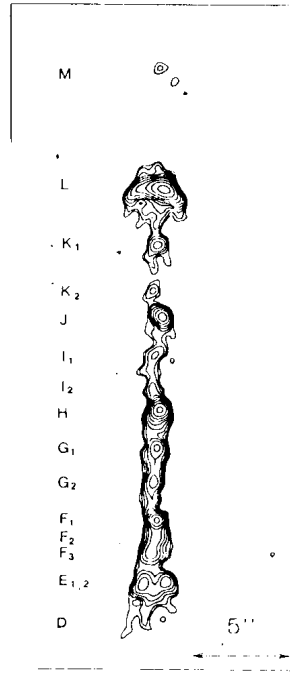

**Fig. 1.** A contour plot of the HH 111 jet, based on an H$\alpha$ image obtained at the NTT in $0.7''$ seeing, and deconvolved to a resolution of $0.39''$. From Reipurth, Raga, Heathcote (1992).

### 2.1 High Resolution Imaging

HH complexes are often large (up to 10 arcmin or more) and they have a wealth of structure at the sub-arcsecond level. It would therefore be very attractive if the VLT was equipped with an imaging instrument like the Wide Field Direct Visual Camera (Wampler 1994), that fully exploit the imaging capabilities of an 8m telescope and take advantage of the image deconvolution techniques which have become increasingly advanced in recent years. Many short exposures can be combined such that the poorest images are used to improve the S/N, while the best ones are defining the resolution of the final image (e.g. Lucy 1992). It is conceivable that resolutions of $0.1''$ can eventually be achieved this way. HST images already obtained of HH complexes provide this resolution, but experience shows that even with 4-5 hour exposures, the 2.4m HST is limited to study only the very brightest HH jets.

High resolution and good S/N will allow to address the as yet unresolved question of the structure of jet knots. If knots are due to time variability in the driving source, they should show the morphology of tiny bow shocks. Another question is whether the large diffuse structures we see in leading working surfaces are already fully resolved, or they have further sub-structure, an important issue for shock models. Further, it is known that working surfaces are divided into bow shocks and Mach disks, with different emission characteristics. It is not known if jet knots are similarly sub-divided, an issue also connected to their mechanism of formation. Finally, HH jets generally show large proper motions, around $1''$-$2''$ per decade. Hitherto, measured proper motions relate to bulk structures, and nothing is known about the detailed flow behaviour on scales that reveal the detailed interaction with the ambient medium.

### 2.2  2-Dimensional Spectroscopy

No high-resolution Fabry-Perot spectrometer is currently being considered for the VLT. But the area spectroscopy (ARGUS) mode of the Multi-Fibre Area Spectrograph (MFAS) allows at least 670 fibres with $0.2''$ or $0.7''$ spatial sampling within 5 or 18 square arcsec fields, with a resolving power of up to 30.000. This is ideally suited for studying the detailed kinematics of selected areas of HH flows. Firstly, accurate radial velocities are required, in combination with proper motions, to determine space motions of HH jets. Secondly, it is of great interest to search for a kinematic signature of the entrainment that takes place when a jet penetrates its environment, since entrainment is likely to have a major impact on the dynamics of the flow. Thirdly, a detailed kinematic study of HH jets can decide if the pronounced wiggling seen in many HH jets is due to precession of the source, flow along a curved tube, or meandering through an inhomogeneous environment, since each of these scenarios predicts a different position-velocity diagram.

### 2.3  Very High Resolution Spectroscopy

The best collimated HH jets lend themselves well to long slit spectroscopy. The UV-Visual Echelle Spectrograph (UVES) permits spectra with resolving powers of the order of 100.000. This resolution corresponds more or less to the thermal widths of the C, N, O, S etc. lines for a gas at 10000 K. With this resolution one would thus get "everything that there is to know" about the radial velocity structure of HH flows. This will allow to address a number of questions: First, are flows in HH jets laminar or turbulent, or do they evolve from one mode into the other? Second, models make specific predictions for the line widths and radial velocities of various emission lines in the recombination region behind a shock wave, which can be tested by very high resolution spectroscopy. Third, the origin of jets is not understood, and very high resolution spectroscopy of residual jets very close to T Tauri stars, as pioneered by Solf (1989), may give clues. Indeed, it is possible that vestiges of the jet phenomenon may be commonly observed in CTTSs if sufficently high spatial and spectral resolution is applied.

### 2.4 Infrared Imaging and Spectroscopy

Many HH jets have in recent years been found to emit strongly in molecular hydrogen and infrared [FeII] lines. The Infrared Spectrometer and Array Camera (ISAAC), the High-Resolution Near-Infrared Camera (CONICA), the Mid Infrared Imager/Spectrometer (MIIS), and the VLT High-Resolution IR Echelle Spectrometer (CRIRES) will be important for this new and rapidly developing field. Some current issues are: First, molecular hydrogen traces the slower and weaker parts of the shocks, and therefore are expected to show different structures than optical lines; when observed with sufficent spatial resolution, they can provide sensitive tests of theoretical models. Second, the youngest objects are most deeply embedded, and it is to be expected that a large population of infrared, optically obscured HH jets await discovery, as exemplified by the recent discovery of HH 212 (Zinnecker et al. 1995). These very young, infrared jets are likely to provide new insights into the origin of jets and their interaction with the ambient medium. Third, infrared lines permit determination of excitation temperatures, electron densities and extinction of HH flows (e.g. Gredel et al. 1992).

### 2.5 VLTI

Extended objects like HH jets will require very good coverage of the (u,v) plane. This might require observations spread out over months. However, with milli-arcsec resolution the large proper motions of jets become a problem: an HH jet would displace itself about one resolution element within 2 or 3 days. In its currently proposed configuration, VLTI is therefore unlikely to have a major impact on the study of HH jets.

## 3 Pre-Main Sequence Binaries

Most stars are members of binary systems. Thus, to understand the process of star formation we must understand the formation of binary stars. As a first step in this direction, a number of recent studies have focused on the properties of binaries which have not yet reached the main sequence (e.g. Reipurth & Zinnecker 1993, Leinert et al. 1993, Ghez et al. 1993). A major review has recently appeared by Mathieu (1994). A principal result deduced in these investigations is that binarity is higher among young stars than among the main sequence stars studied by Duquennoy & Mayor (1991). This could be because the distribution of separations of components is different for younger and older stars, even though the total binary frequency is constant, and that there is an excess only in the separation range sampled by the observations. Or perhaps essentially all stars are born in binary or multiple systems, which then subsequently are disrupted, possibly through close encounters with other stars. In either case, it is clear that young binaries undergo dynamical evolution towards the main sequence. It will be important in the coming years to do unbiased surveys for faint sub-arcsecond companions to young stars in the visible and the infrared, and the VLT will be able to produce outstanding results in this field. Speckle observations will push the resolution limit even further. But one should remember that while for diffraction-limited speckle observations the resolution limit will improve going from a 3.6m to an 8m telescope, there is no similar gain in limiting magnitude.

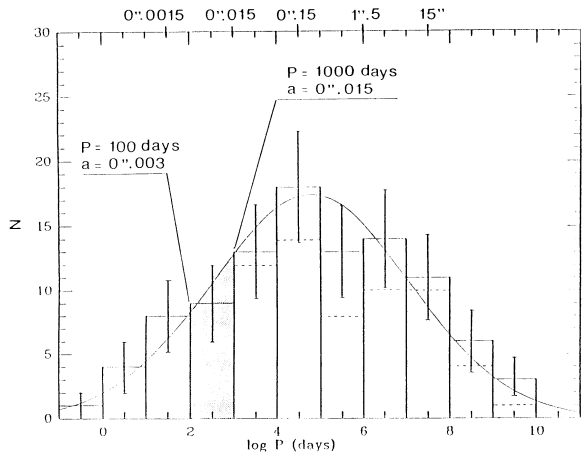

**Fig. 2.** The period distribution for nearby G dwarfs, from Duquennoy & Mayor (1991). The periods in days have been converted to semi-major axes in arcsec as described in the text. Practical observing limits with VLTI have been marked, the lower limit is $a = 0.003''$ ($P = 100$ days) and the upper limit is $P = 1000$ days ($a = 0.015''$). This range covers about 10% of all binaries in nearby (130 pc) star forming regions.

The really unique contributions to the study of pre-main sequence binaries will be made by the VLTI. The question is whether VLTI allows us to resolve the orbits of spectroscopic PMS binaries. If so, an orbit determination gives us the sum of the component masses, while the spectroscopic data give us the mass ratio (or the mass function if the system is only single-lined). Together the data would give us the first precise mass determinations of low-mass pre-main sequence stars, in much the same way it has been done for nearby low-mass binaries (e.g. Mariotti et al. 1990). This would for the first time allow a meaningful observational calibration of theoretical evolutionary tracks before the main sequence. If we assume observations to be done at $\lambda$ 5500 Å with a VLTI baseline of 150 m, we derive $\lambda/D = 0.00076'' \approx 1$ milliarcsec. The distance of the nearest star forming complexes is about 130 pc, which we will adopt in the following, and let us assume that the binary components have $m_1 = m_2 = 0.5~M_\odot$. We shall additionally assume that the period distribution of low-mass PMS binaries is not too different from the distribution of G-dwarfs determined by Duquennoy & Mayor (1991), an assumption that is not necessarily correct, but which is the best one can do before an actual determination is done. Figure 2 shows the G-dwarf period distribution, with periods converted into semi-major axes in arcseconds. If we assume that 3 milliarcsec (corresponding to a period of 100 days) is a lower limit for an orbit determination, and that systems with periods longer than 1000 days are difficult to observe in practice, then we find that about 10% of all PMS binaries have separations large enough for an orbit determination and periods short enough to be observable over a full orbital revolution. It is therefore of particular interest to find and study PMS binaries in the period interval from 100 to 1000 days.

However, it is important to recall that distances to star forming regions are not very precisely determined. Even if we could claim to know these distances to as well as 10%, this translates via Keplers law into a 30% uncertainty in the masses, since $[a(1+0.1)]^3 \approx a^3 + 0.3a^3 + \cdots$. It is therefore clear that an integral part of a PMS binary mass determination must be a *distance* determination, e.g. by measuring parallaxes via narrow-angle VLTI astrometry, as discussed elsewhere in this volume.

## 4 Circumstellar Disks

The presence of circumstellar disks around young stars is inferred from indirect clues rather than direct observations. The most tantalizing arguments for the presence of disks rely on *models* of (a) forbidden line profiles, (b) spectral energy distributions, and (c) polarization patterns.

Search for direct evidence of disks around young stellar objects is an active research area, and observations in various wavelength ranges uncovered several types of condensations and/or disk-like structures in the immediate vicinity of young stars. For example,

- Near-infrared adaptive optics observations at the ESO 3.6m telescope showed that a 400AU disk-like structure surrounds the FU Orionis binary star Z CMa (Malbet et al. 1993).
- Millimeter and submillimeter range interferometry of the T Tauri star HL Tau showed that the star is embedded in molecular condensations on various scales. Kinematic evidence for entrainment of molecular gas by the jet is apparent on the scale of a few arcsecond (Cabrit et al. 1994).
- Also in the millimeter range, the T Tauri binary star GG Tau was found to be surrounded by a circumbinary molecular ring (Dutrey et al. 1994).

These examples attest to the large variety of circumstellar environnments that are associated with young stars but give little evidence so far for the Keplerian accretion disks that are thought to be responsible for the FU Orionis outbursts and the strong activity of CTTSs. Clearly, a much higher spatial resolution than currently available is necessary to probe the structure of these disks and to test current ideas about the physical processes at work in these objects.

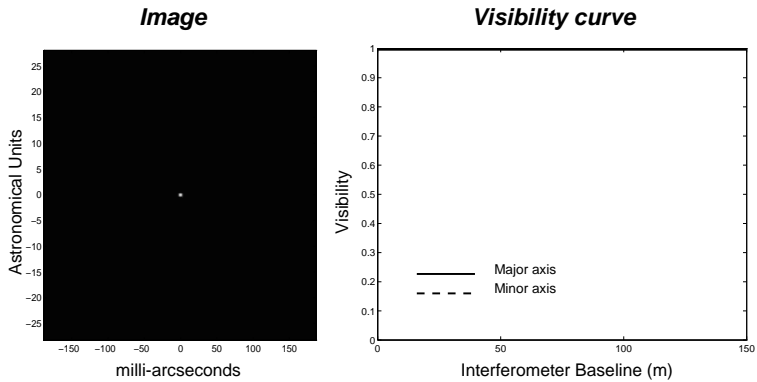

**Fig. 3 a.** $0.5\mu$m direct image (left) and visibility (right) of a typical T Tauri accretion disk. The visibility is constant and equal to 1 because the circumstellar environment is unresolved.

In order to compare expected accretion disk properties to expected performances of the VLT and VLTI, we computed synthetic images and visibilities of the *thermal* radiation emitted by a typical T Tauri disk model. Figures 3a to 3c display disk images and visibilities at 0.5, 2.2, and 10 $\mu$m and Table 1 summarizes the 2.2 $\mu$m performances at (a) one VLT telescope equipped with adaptive optics and used as an imaging device, and (b) the VISA interferometer with three 1.8m telescopes and up to 150m baseline. The visibility is a way to quickly see if an object is resolved with an interferometer and two or three different (u,v) plane configuration is enough to understand the global morphology of a disk (inclination, size,...).

A first conclusion that can be drawn from comparing Figure 3 and Table 1 is the usefulness of the VISA interferometer for probing the physical properties of the inner accretion disk. Indeed, the 8m telescope 2.2$\mu$m resolution with adaptive optics *full* correction – an optimistic estimate – is about 10 AU at the distance of the closest star-forming regions while VISA allows sub-AU resolution at the same distance. Figure 3b demonstrates that a few interferometric measurements in the (u,v) plane will allow one to determine roughly the geometry of 2.2$\mu$m emission by simple visibility modeling.

A second obvious conclusion is the usefulness of 10$\mu$m measurements for studying the circumstellar environment by direct imaging. This is the result of two conspiring factors: (a) 8m telescopes are diffraction-limited at this wavelength, and (b) because of the low surface temperature of disks surrounding young stars, we see the entire disk at wavelengths close to or larger than $10\mu m$. An additional bonus not taken into account here is the low extinction in the mid-IR, which allows one to study the dense star-forming cores. We thus wish to emphasize the strong need for modern mid-infrared arrays at the VLT if this facility is to be competitive in the field of star formation studies.

**Table 1.** Observations of T Tauri Disks: VLT vs. VISA

|  | VLT with *maximum* adaptive optics correction (1 x 8m) | VISA (3x1.8m telescopes) |
|---|---|---|
| Technique | direct imaging (image deconvolution in principle not needed) | visibility measurement (model constraining); ultimately, image reconstruction |
| Field of view | 30"-90" | 2" |
| Resolution @ 2.2 μm (in mas) | 60 | 5 (for 100m baseline) |
| Resolution in AU (d=150pc) | 9 | 0.75 |
| Limiting magnitude @ 2.2μm | 13 (limited by AO ref. star) | 8 - 12 (fringe tracker limited) |
| On source sensitivity | 25 | 8 - 12 |

Such mid-IR detectors should also be made available for the VISA interferometer, as Table 2 demonstrates. There, we show that even in a very conservative estimate of limiting magnitude, which corresponds to a case of observations with no fringe tracker, the expected disk model flux is larger than VISA's limiting flux up to the near-IR. Given a sufficient coverage of the (u,v) plane, one could thus map the disk temperature structure and distinguish between the various physical processes that may be responsible for disk thermal emission (kinematic viscosity, magnetic viscosity, reprocessing of stellar light, etc).

**Table 2.** Observations of T Tauri accretion disks with VISA: a conservative estimate (lim. mag. K=8)

| Wavelength (μm) | VISA limiting flux (Jy) | Disk model flux (Jy) |
|---|---|---|
| 2.2 | 0.8 | 3.8 |
| 5 | 0.4 | 4.8 |
| 10 | 3.2 | 4.5 |
| 20 | 11 | 3.8 |

*Notes to Table 2.* The following assumptions were made in deriving the above values.
- 3 x 1.8m telescopes
- Detector or background noise-limited case
- S/N=50
- Total optical efficiency 10%
- Visibility instrumental losses 30%
- Classical accretion disk model with mass-accretion rate of $10^{-6}\,M_\odot/yr$

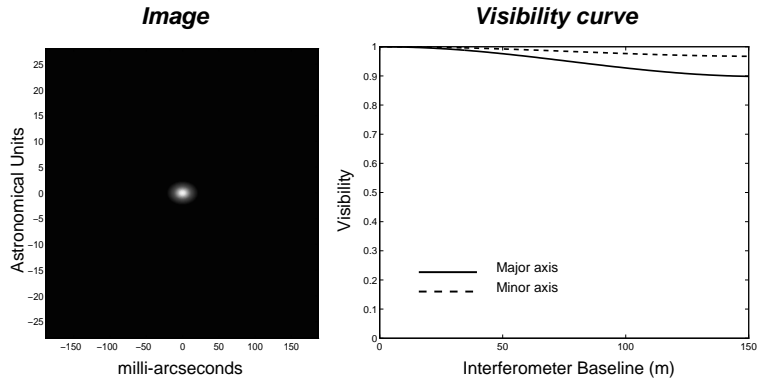

**Fig. 3 b.** $2.2\mu$m direct image (left) and visibility (right) of a typical T Tauri accretion disk. The decrease at long baseline shows the disk is resolved. A visibility accuracy of 1% will easily permit to measure a visibility of 93% with 100-m baseline.

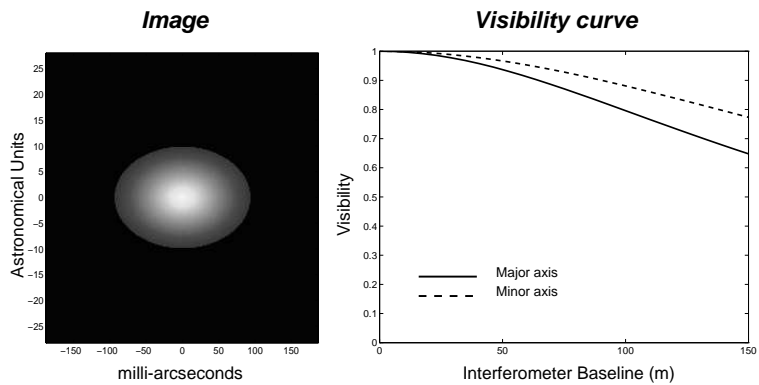

**Fig. 3 c.** $10\mu$m direct image (left) and visibility (right) of a typical T Tauri accretion disk. The visibility curve shows that the disk is easily resolved.

In summary, we emphasize that any 8-10m telescope equipped with adaptive optics will allow for significant progress in understanding of circumstellar environments. However, a VISA-type interferometer is needed for qualitative breakthrough in studies of planetary system formation and evolution. The $10\mu$m window provides the best resolution/flux compromise for studying disks around YSOs, and VISA (better science version) has the potential performances needed for detecting T Tauri and FU Orionis disks in the near IR (strong deviations from axisymmetry are expected in the disk). One should note that the Palo-mar ASEPS-0 Testbed interferometer (prototype for the Keck interferometer), will soon be operational and is expected to reach K= 6-7 mag. Therefore, the so-called better science version is the minimum configuration that should be considered for a competitive VISA, even if its construction is delayed by one or two years.

## 5 Surface Properties of Young Stars

Progress on surface properties of bright stars has been very quick in recent years, due to the conjunction of long-term synoptic observations and high-resolution spectroscopy. Young stars being quite faint, large collecting areas are necessary to accurately derive their surface properties. We focus in this section on two topics of current interest in which obvious progress can be achieved with the UVES spectrograph.

### 5.1 Magnetic Field Strength and Topology in Young Stars

On the basis of comparative studies with active magnetic main-sequence stars, one expects magnetic field strengths of 1-2 kG in T Tauri stars. Actual measurements are difficult; Zeeman broadening of photospheric lines is given by $\Delta\lambda = 5.7\,10^{-10} g \lambda^2 B$, where $g$ is the effective Landi factor, $\lambda$ is in Angstroms, $B$ in kG, e.g., $\Delta\lambda$= 0.05Å for $g$=2.5, $B$=1kG, $\lambda$=6000Å. So, far, there is only one published measurement, which gives B = $1.0 \pm 0.5$kG for WTTS TAP 35 (Basri and Marcy 1990).

A current observing method for magnetic field strengths involves correlating the incident spectrum with numerical masks containing lines with similar Landi factors, thus giving high S/N correlation peaks which can be calibrated to give B (Queloz, Babel, Mayor, in preparation). Performance on existing 2m-telescope are R=40000, S/N=20, for a 1h integration time, allowing for detecting B $\approx$ 2kG in a star with V=11.

Expected UVES performance is about the same but for V=16, so that T Tauri stars can be studied. Measurement of B in a large sample of WTTSs (V $\leq$ 14) with a time resolution of 1h will allow to study (a) the distribution of magnetic field on the stellar surface, (b) the origin of magnetic field (dynamo vs. fossil field) by looking for correlations between B and rotational period, (c) the evolution of B during pms phase, and, (d) in conjunction with Doppler imaging (see below), the correlations between cold spots and chromospheric network.

Measurement of B in FU Orionis spectra, a difficult but potentially quite rewarding experiment, will allow one to study the distribution of magnetic field on the disk surface and to test the magnetodynamic origin of FU Ori winds and jets.

### 5.2 Doppler Imaging of Young Stars

For a V=11 star, i.e., for the brightest T Tauri stars, spectrograms with R=40000, S/N=100, obtained in t=1h on an existing 2m-class telescope allow for Doppler-mapping spots with diameter 15% of the star's diameter or larger, and a spot-to-photosphere temperature contrast larger than 800K (cf. Joncour et al. 1994).

A possible project with UVES will be to Doppler-image the stellar surface of relatively bright CTTSs to resolve either the small hot spots expected if accretion is channeled by the magnetic field lines or the (non-axisymmetric) boundary layer. With this technique, we will thus be able to study the nature of accretion onto T Tauri stars, a most timely and interesting problem in star formation research. Again, a word of caution is in order, since the HIRES spectrograph of the Keck telescope allows our Californian colleagues to solve this problem today.

**Acknowledgements**

We are grateful to Jérôme Bouvier, Christian Perrier, Alex Raga and Hans Zinnecker for valuable comments and discussions.

## References

Basri, G., Marcy, G. (1990). Limits on the Magnetic Flux on a Pre-Main Sequence Star. In The Sun and Cool Stars: Activity, Magnetism, Dynamos (IAU Coll. No.130). Helsinki: Touminen.
Cabrit, S., Guilloteau, S., André, P., Bertout, C., Montmerle, T. (1994). Astron. Astrophys. submitted.
Duquennoy, A., Mayor, M. (1991): Astron. Astrophys. 248, 485
Dutrey, A., Guilloteau, S., Simon, M. (1994). Astron. Astrophys. 286, 149
Ghez, A.M., Neugebauer, G., Matthews, K. (1993): Astron.J. 106, 2005
Gredel, R., Reipurth, B., Heathcote, S. (1992): Astron. Astrophys. 266,439
Joncour, I., Bertout, C., Mńard, F. (1994). Astron. Astrophys. 285, L25-L28.
Leinert, Ch., Zinnecker, H., Weitzel, N., Christou, J., Ridgway, S.T., Jameson, R., Haas, M., Lenzen, R. (1993): Astron. Astrophys. 278, 129
Lucy, L.B. (1992): Astron.J. 104,1260
Malbet, F., Rigaut, F., Bertout, C., Léna, P. (1993). Astron. Astrophys. 271, L9
Mariotti, J.-M., Perrier, C., Duquennoy, A., Duhoux, P. (1990): Astron. Astrophys. 230, 77
Mathieu, R. (1994): Ann. Rev. Astron. Astrophys. 32, 465
Raga, A.C., Cabrit, S. (1993): Astron. Astrophys. 278,267
Reipurth, B. (1989): Nature 340, 42
Reipurth, B., Heathcote, S. (1993): in *Astrophysical Jets*, STScI Symp. Series vol.6, eds. D.Burgarella, M. Livio, C.P.O'Dea, p.35
Reipurth, B., Zinnecker, H. (1993): Astron. Astrophys. 278,81
Reipurth, B., Raga, A.C., Heathcote, S. (1992): Astrophys.J. 392,145
Solf, J. (1989): in ESO Workshop on *Low Mass Star Formation and Pre-Main Sequence Objects*, ed. Bo Reipurth, p.399

Wampler, E.J. (1994): in *Instruments for the ESO VLT*, ed. A.F.M. Moorwood, ESO, p.55

Zinnecker, H., McCaughrean, M., Rayner, J. (1995): in press